\documentclass[twocolumn,showpacs,aps,prd,floatfix]{revtex4}
\usepackage{amssymb}
\usepackage{amsmath}
\usepackage{color}
\usepackage{graphicx}
\usepackage{dcolumn}
\usepackage{epsfig}
\usepackage{colordvi}
\usepackage{color}
\usepackage{hhline}

\usepackage{dcolumn}
\usepackage{bm}
\usepackage{subfigure}
\usepackage{multirow}
\usepackage{braket}
\usepackage{hyperref}

\RequirePackage{xspace}

\newcommand{\BaBarYear}       {15}
\newcommand{\BaBarNumber}     {009}
\newcommand{\SLACPubNumber} {16505}
\newcommand{\BaBarType}      {PUB}  

\input{babarsym}

\long\def\inst#1{\par\nobreak\kern 4pt\nobreak
   {\it #1}\par\vskip 10pt plus 3pt minus 3pt}

\newcolumntype{,}[1]{D{,}{\pm}{#1}}
\begin{document}


\begin{flushleft}
\babar-\BaBarType-\BaBarYear/\BaBarNumber \\
SLAC-PUB-\SLACPubNumber 
\end{flushleft}

\title{\large \bf
    \boldmath
        Measurement of the neutral $D$ meson mixing parameters in a time-dependent amplitude analysis of the $\Dz\to\pip\pim\piz$ decay}


\newcommand{\xfitcent}{2.08}
\newcommand{\yfitcent}{0.14}
\newcommand{\xcent}{1.50}
\newcommand{\ycent}{0.19}
\newcommand{\xcentf}{1.5}
\newcommand{\ycentf}{0.2}
\newcommand{\tcent}{410.2}

\newcommand{\xstat}{1.17}
\newcommand{\ystat}{0.89}
\newcommand{\xstatf}{1.2}
\newcommand{\ystatf}{0.9}
\newcommand{\tstat}{3.8}

\newcommand{\xsys}{0.56}
\newcommand{\ysys}{0.46}
\newcommand{\xsysf}{0.6}
\newcommand{\ysysf}{0.5}
\newcommand{\tsys}{?.?}

\newcommand{\xfinal}{\xcent \pm \xstat \pm \xsys}
\newcommand{\xfinalf}{\xcentf \pm \xstatf \pm \xsysf}
\newcommand{\xfinalStat}{\xcent \pm \xstat}
\newcommand{\xfitStat}{\xfitcent \pm \xstat}
\newcommand{\yfinal}{\ycent \pm \ystat \pm \ysys}
\newcommand{\yfinalf}{\ycentf \pm \ystatf \pm \ysysf}
\newcommand{\yfinalStat}{\ycent \pm \ystat}
\newcommand{\yfitStat}{\yfitcent \pm \ystat}
\newcommand{\tfinal}{\tcent \pm \tstat \pm \tsys}
\newcommand{\tfinalStat}{\tcent \pm \tstat}

\newcommand{\putat}[3]{\begin{picture}(0,0)(0,0)\put(#1,#2){#3}\end{picture}}

%
%
\author{J.~P.~Lees}
\author{V.~Poireau}
\author{V.~Tisserand}
\affiliation{Laboratoire d'Annecy-le-Vieux de Physique des Particules (LAPP), Universit\'e de Savoie, CNRS/IN2P3,  F-74941 Annecy-Le-Vieux, France}
\author{E.~Grauges}
\affiliation{Universitat de Barcelona, Facultat de Fisica, Departament ECM, E-08028 Barcelona, Spain }
\author{A.~Palano}
\affiliation{INFN Sezione di Bari and Dipartimento di Fisica, Universit\`a di Bari, I-70126 Bari, Italy }
\author{G.~Eigen}
\affiliation{University of Bergen, Institute of Physics, N-5007 Bergen, Norway }
\author{D.~N.~Brown}
\author{Yu.~G.~Kolomensky}
\affiliation{Lawrence Berkeley National Laboratory and University of California, Berkeley, California 94720, USA }
\author{H.~Koch}
\author{T.~Schroeder}
\affiliation{Ruhr Universit\"at Bochum, Institut f\"ur Experimentalphysik 1, D-44780 Bochum, Germany }
% ...
\collaboration{The \babar\ Collaboration}
\noaffiliation

\begin{abstract}
We perform the first measurement on the $\Dz - \Dzb$ mixing parameters using a time-dependent amplitude analysis of 
the 
decay $\Dz\to\pi^+\pi^-\pi^0$. 
The data were recorded with the \babar\ detector at center-of-mass 
energies at and near the $\Upsilon(4S)$ resonance, and correspond to an integrated luminosity of 
approximately $468.1\invfb$. The neutral $D$ meson candidates are selected from 
$D^{*}(2010)^+\to D^0 \pi_s^+$ decays where the flavor at the production is identified by the charge of the 
low momentum pion, $\pi_s^+$. The measured mixing parameters are $x = (\xfinalf) \%$ and $y = (\yfinalf) \%$, where the 
quoted uncertainties are statistical and systematic, respectively. 
\end{abstract}



\maketitle                                                                             

\setcounter{footnote}{0}

\newcommand{\ie}{\textit{i}.\textit{e}.}

\section{Introduction}
\label{sec:Introduction}

The first evidence for $\Dz-\Dzb$ mixing, which had been sought for more than two decades since it 
was first predicted~\cite{PhysRevD.12.2744}, was obtained by \babar\ ~\cite{PhysRevLett.98.211802} 
and Belle~\cite{PhysRevLett.98.211803} in 2007. 
These results were rapidly confirmed by 
CDF~\cite{PhysRevLett.100.121802}. 
The techniques utilized in those analyses
and  more recent, much higher statistics LHCb analyses 
\cite{Aaij:2012nva,Aaij:2013wda,Aaij:2016rhq}
do {\em not} directly measure
the normalized mass and the
width differences of the neutral $ D $
eigenstates, $ x $ and $ y $.
In contrast,
a time-dependent amplitude analysis of  the Dalitz-plot (DP)  of neutral $ D $
mesons decaying into self-conjugate final states provides direct measurements
of both these parameters.
This technique was introduced using $ D^0 \to K^0_S \pi^- \pi^+ $
decays by the CLEO collaboration \cite{Asner:2005sz},
and the first measurement by the Belle Collaboration \cite{Abe:2007rd}
provided  stringent constraints on the mixing parameters.
More recent measurements with this final
state by the {\babar} and Belle collaborations
\cite{delAmoSanchez:2010xz,Peng:2014oda} contribute significantly
to the  Heavy Flavor Averaging Group (HFAG) global fits that determine 
world average mixing and \CP violation parameter values~\cite{hfag2014}. 

This paper reports the first
measurement of mixing parameters from a time-dependent
amplitude analysis of the singly Cabibbo-suppressed decay $\Dz\to\pipi\piz$. 
The inclusion of 
charge conjugate reactions is implied throughout this paper.
No measurement of \CP violation is attempted as the data set 
lacks sufficient sensitivity to be interesting. 
 The $\Dz$ candidates are selected 
from $D^{*}(2010)^+\to\Dz\pi^+_s$ decays where the $\Dz$ 
 flavor at production is identified by the charge 
of the slow pion, $ \pi^+_s $. 

The $\Dz$ and $\Dzb$ meson  
flavor eigenstates 
evolve and decay as mixtures of the weak Hamiltonian eigenstates $D_1$ 
and $D_2$ with masses and widths $m_1, \Gamma_1$ and $m_2, \Gamma_2$, 
respectively. 
The
mass eigenstates can be expressed as superpositions of the flavor eigenstates, 
$\Ket{D_{1,2}} = p\Ket{\Dz} \pm q\Ket{\Dzb}$ where the complex
coefficients $p$ and $q$ satisfy 
$\left|p\right|^2 + \left|q\right|^2 = 1$. 
The mixing parameters are defined as normalized mass and
width differences, $x \equiv (m_1-m_2)/\Gamma_D$ 
and $y \equiv (\Gamma_1-\Gamma_2)/2\Gamma_D$.
Here, $ \Gamma_D $ is  the average decay width, $\Gamma_D \equiv (\Gamma_1+\Gamma_2)/2$. 
These mixing 
parameters appear in the expression for the decay rate at 
each  point $\left(s_+,s_-\right)$
in the $\Dz$ decay Dalitz-plot at the decay 
time $t$, where $s_\pm \equiv m^2(\pi^\pm\pi^0)$.
For a charm meson tagged at $t=0$ as a $\Dz$,  
the decay rate is proportional to

\begin{align}
    \left|{\cal M}(\Dz)\right|^2 &\propto \frac{1}{2} e^{-\Gamma_D t}  \left\{\left|A_f\right|^2\left[\cosh\left(y\Gamma_D t\right)+\cos\left(x\Gamma_D t\right)\right] \right.\nonumber\\
 & +\left|\frac{q}{p}\overline{A}_f\right|^2\left[\cosh(y\Gamma_D t)-\cos(x\Gamma_D t)\right] \nonumber\\
 & - 2\left[ {\rm{Re}}\left(\frac{q}{p}A_f^*\overline{A}_f\right) \sinh(y \Gamma_D t) \right.\nonumber\\
 & \left. \left. - {\rm{Im}}\left(\frac{q}{p}A_f^*\overline{A}_f\right) \sin(x \Gamma_D t) \right]\right\}\,,
\label{eq:mixrate}
\end{align}
where $f$ represents the $\pip\pim\piz$ final state 
that is commonly accessible
to decays of both flavor eigenstates, 
and
$A_f$ 
and $\overline{A}_f$ are the decay amplitudes for \Dz and \Dzb to final state $f$. 
The amplitudes are functions 
of position in the DP and are defined in our description of the fitting model in Sec.~\ref{sec:fitmodel} 
Eq.~(\ref{eq:defamp}). 
In Eq.~(\ref{eq:mixrate}), the first term is 
the direct decay rate to the final state $f$ and is always the dominant
term for sufficiently small decay times. 
The second term corresponds to mixing. 
Initially,
the $\cosh(y\Gamma_D t)$ and $\cos(x\Gamma_D t)$
contributions to this term cancel, but over time the $\cosh(y\Gamma_D t)$ contribution can become dominant.
The third term is the interference term. It depends explicitly on the real and imaginary parts of
$A_f^*\overline{A}_f$ and on the real and imaginary parts
of $q/p$. As for the mixing rate, the interference rate is intially zero, but it can become
important at later decay times. 
The variation of the total decay rate from purely exponential
depends on the relative strengths of the direct and mixing amplitudes, their relative phases, the mixing
parameters $x$ and $y$, and on the magnitude and phase of $q/p$. 
HFAG reports the world averages to be 
$ x = ( 0.49 ^{+0.14}_{-0.15} ) \% $ and
$ y = ( 0.61 \pm 0.08 ) \% $ assuming  no  \CP\ violation~\cite{hfag2014}. 

In this time-dependent
amplitude analysis of the DP, we  measure $ x $, $ y $, 
$ \tau_D \equiv 1/\Gamma_D$, and resonance parameters of the decay model. 
At the level of precision of this measurement, \CP\ violation can be neglected. 
Direct \CP\ violation in this channel is well constrained~\cite{pdg},
and indirect \CP\ violation due to $q/p \neq 1$ is also very small, as 
reported by HFAG~\cite{hfag2014}.
We assume no 
\CP\ violation, \ie, 
$q/p = 1$, and $\overline{A}_f(s_+,s_-) = A_f(s_-,s_+)$.

This paper is organized as follows: Section~\ref{sec:detector} discusses the \babar\ detector 
and the data used in this analysis.
Section~\ref{sec:evtsel} describes the event selection. 
Section~\ref{sec:fitstrategy} presents the model used to describe the amplitudes in the DP and the 
fit to the data. Section~\ref{sec:systematics} discusses and quantifies the sources of systematic uncertainty. 
Finally, the results are summarized in Sec.~\ref{sec:conclusion}.


\section{\boldmath The \babar\ detector and data}
\label{sec:detector}
This analysis is based on a data sample corresponding to an integrated luminosity of 
approximately $468.1$~\invfb recorded at, and $40 \mev$ below, the $\Upsilon\left(4S\right)$ 
resonance by the \babar\ detector at the \pep2\ asymmetric energy \epem\ collider~\cite{Lees2013203}. 
The \babar\ detector is described in detail elsewhere~\cite{ref:babar, ref:nim_update}.
Charged particles are measured with a combination 
of a 40-layer cylindrical drift chamber (DCH) and a 5-layer double-sided silicon vertex tracker (SVT), 
both operating within the $1.5$ T magnetic field of a superconducting solenoid. Information from a 
ring-imaging Cherenkov detector is combined with specific ionization $(dE/dx)$ measurements from 
the SVT and DCH to identify charged kaon and pion candidates. Electrons are identified, and photons 
measured, with a CsI(Tl) electromagnetic calorimeter. The return yoke of the superconducting coil is 
instrumented with tracking chambers for the identification of muons.

\section{Event selection}
\label{sec:evtsel}

We reconstruct $D^{*+}\rightarrow D^0 \pi_s^+$ decays coming from $\epem\to\ccbar$ 
in the 
channel $\Dz\to \pi^+\pi^-\pi^0$. 
$\Dstarp$ candidates from $B$-meson decays are disregarded due to
high background level. 
The pion from the $D^{*+}$ decay is called the 
``slow pion'' (denoted $\pi_s^+$) because of the limited phase space available. The mass difference of 
the reconstructed $D^{*+}$ and $D^{0}$ is defined as 
$\Delta m \equiv m\left(\pi^+\pi^-\pi^0\pi_s^+\right) -  m\left(\pi^+\pi^-\pi^0\right)$. Many of the selection criteria 
and background veto algorithms discussed below are based upon previous \babar\ 
analyses~\cite{PhysRevD.74.091102, PhysRevLett.99.251801}. 

To select well-measured slow pions,
 we require that the $\pi_s^+$ tracks have at least $10$ hits measured
in the DCH; and we reduce backgrounds from other non-pion tracks by requiring that the $dE/dx$ values 
reported by the SVT and DCH be consistent with the pion hypothesis. The Dalitz decay 
$\pi^0\rightarrow \gamma e^+ e^-$ produces background when we misidentify the $e^{+}$ as a $\pi_s^+$.  
We reduce such background by trying to reconstruct 
an $\epem$ pair using the candidate $\pi_s^+$ track as the $\ep$
and combine it with a $\gamma$.
If the $e^+e^-$ vertex is within the SVT volume and the invariant mass is in the range 
$115 < m\left(\gamma e^+ e^-\right) < 155 \mev$, then the event is rejected. Real photon conversions 
in the detector material are another source of background in which electrons can be misidentified as slow pions. 
To identify such conversions, we first create a candidate $e^+e^-$ pair using the slow pion candidate and an 
identified electron, 
and perform a least-squares fit. 
The event 
is rejected if the invariant mass of the putative pair is less than $60 \mev$ and the constrained vertex position is 
within the SVT tracking volume. 

We require that the $D^0$ and $\pi_s^+$ candidates originate 
from a common vertex, and that
the $D^{*+}$ candidate  
originates from the $e^+ e^-$ interaction region (beam spot). 
A kinematic fit to the entire decay chain is performed
with geometric constraints at each decay vertex. In addition, 
the $\gamma\gamma$ and $\pi^+\pi^-\pi^0$ invariant masses are constrained 
to be 
the nominal $\pi^0$ and $D^0$ masses, respectively~\cite{pdg}. 
The  
$\chi^2$ probability 
of the $D^{*+}$ fit 
must be at least 0.1\%. 
About 15\% of events with at least one candidate satisfying all selection
criteria (other than the final $ D^0 $ mass
and $ Delta m $ cuts described below) have at least two such candidates.
In these events,
we select the candidate with 
the smallest $\chi^2$ value.

To suppress 
misidentifications from low-momentum neutral pions, we require the laboratory momentum of the $\pi^0$ candidate 
to be greater than 350 \mev. The reconstructed $\Dz$ proper decay time $t$,
 obtained from our kinematic fit,  must be 
within the time window $-2< t <3$~ps 
and have an uncertainty $\sigma_t < 0.8$~ps. Combinatorial and $B$ meson decay 
background is removed by requiring 
$p^*(D^{0}) > 2.8 \gev$, where $p^*$ is the momentum measured in the \epem center-of-mass frame for the 
event. The reconstructed $D^0$ mass must be within 15 \mev of the nominal \Dz mass~\cite{pdg} and the 
reconstructed $\Delta m$ must be within 0.6 \mev of the nominal $D^{*+}$--$D^0$ mass difference~\cite{pdg}. 
After imposing all other event selection requirements as mentioned earlier, 
these $p^*(D^{0})$, $\sigma_t$, $m(\pi^+\pi^-\pi^0)$, and $\Delta m$ criteria were chosen to maximize the  
significance of the signal yield
obtained from a 2D-fit to the $m, \Delta m$ plane of data, 
where the significance was calculated as $S/\sqrt{S+B}$ 
with $S$ and $B$ as the numbers of signal and background events, respectively. 

The signal probability density functions (PDFs) in both $m$ and 
$\Delta m$ are each defined as the sum of two Gaussian functions. The $m(\pi^+\pi^-\pi^0)$ background distribution is 
parameterized by the sum of a linear function and a single Gaussian, which is used to model the 
$\Dz \to \Km \pip \piz$ 
contribution 
when we misidentify the kaon track as a pion. We use a threshold-like function~\cite{argus} to model the $\Delta m$ 
background as a combination of real $ D^0 $ mesons with random slow pion candidates near kinematic threshold. 

For many purposes, we use ``full'' Monte Carlo (MC) simulations
in which each data set 
is roughly the same size as that observed in the 
real data and the background is a mixture of
$\bbbar$, $\ccbar$, $\tautau$ and $\uubar/\ddbar/\ssbar$ events 
scaled to the data luminosity. 
The signal MC component is
generated with  four combinations of $x=\pm1\%,\ y=\pm1\%$. 
We create four samples for each set of
mixing values except $x= y=+1\%$ which has ten samples.

Based upon detailed study of full MC events, we have 
identified four specific misreconstructions 
of the $D^0$ candidate that we can safely remove from the signal region 
without biasing the measured 
parameters. 
The first mis-reconstruction creates a peaking background in the corner 
of the DP when the $\Km$ daughter 
of a $D^0 \to K^- \pi^+$ decay is misidentified as a pion.
To veto these events,
we assign the kaon  
mass hypothesis for the $\pi^+\pi^-$ candidates 
and calculate the $m(K^-\pi^+)$ 
invariant mass. 
We remove more than $95\%$ of these mis-reconstructions by requiring 
$\left|m(K^-\pi^+) - m(D^0)\right| > 20$ \mev. 

The second mis-reconstruction occurs when the $D^0$ signal candidate 
shares one or more tracks with a $D^0 \to K^- \pi^+\pi^0$ decay. 
To veto these decays, we create a list of all $\Dz\to\Km\pip\piz$ 
candidates in the event that satisfy $\left|m(K^-\pi^+\pi^0) - m(D^0)\right| < 20$ \mev, $\left|\Delta m - \Delta m_{\rm PDG}\right| < 3$ \mev, 
and $\chi_{\rm veto}^2 < 1000$, where

\begin{align}
    \chi_{\rm veto}^2 (m, \Delta m) & = \left(\frac{m(K^-\pi^+\pi^0) - m_{\rm{PDG}}(D^0)}{\sigma_m}\right)^2 \nonumber\\
                                    &+ \left(\frac{\Delta m - \Delta m_{\rm{PDG}}}{\sigma_{\Delta m}}\right)^2\,,
\label{eq:vetochi2}
\end{align}
where $m_{\rm PDG}$ denotes the nominal value for the mass taken from Ref.~\cite{pdg} and 
$\sigma_m$ ($\sigma_{\Delta m}$) is the $m$ ($\Delta m$) uncertainty reported by the fit. 
Such an additional veto is applied for the specific case when the $\pi^+\pi^0$ from a $D^0 \to K^- \pi^+\pi^0$ decay 
is paired with a random $\pi^-$ to form a signal candidate. We can eliminate more than $95\%$ of these mis-reconstructions 
by finding the $K^-$ candidate in the event that yields a $m(K^-\pi^+\pi^0)$ invariant mass closest to the nominal 
$D^0$ mass and requiring $\left|m(K^-\pi^+\pi^0) - m(D^0)\right| > 40 \mev$. The background from 
$D^0 \to K^- \pi^+\pi^0$ due to misidentifying the kaon track as a pion falls outside the signal region mass 
window and is negligible. 

The third mis-reconstruction is the peaking background when the 
$\pi^+\pi^-$ pair from a $D^0 \to \KS \pi^+\pi^-$ decay is combined with a random $\pi^0$ to form a signal candidate. 
To veto these events,
we combine the $\pi^+\pi^-$ 
from a $ D^0 \to \pi^+ \pi^- \pi^0 $
candidate with $\KS\to \pi^+\pi^-$ candidates in the same event and 
require $\left|m(\KS \pi^+\pi^-) - m(D^0)\right| > 20$~\mev 
for each. 

The fourth mis-reconstruction is pollution from 
$D^0 \to \KS \pi^0 \to (\pi^+\pi^-)\pi^0$ decay. 
Although a real $ D^0 $ decay, its amplitude does not
interfere with those for ``prompt'' $ D^0 \to \pi^+\pi^- \pi^0 $.
We eliminate $ \sim 99\%$ of these events by removing candidates with  
$475 < m(\pi^+\pi^-) < 505 \mev$. 
The $\KS$ veto also removes other potential backgrounds associated with $\KS$ 
decays.

\begin{figure}[!h]
\centering
    \subfigure{
        \label{fig:massfit-a}
\includegraphics[width=\linewidth]{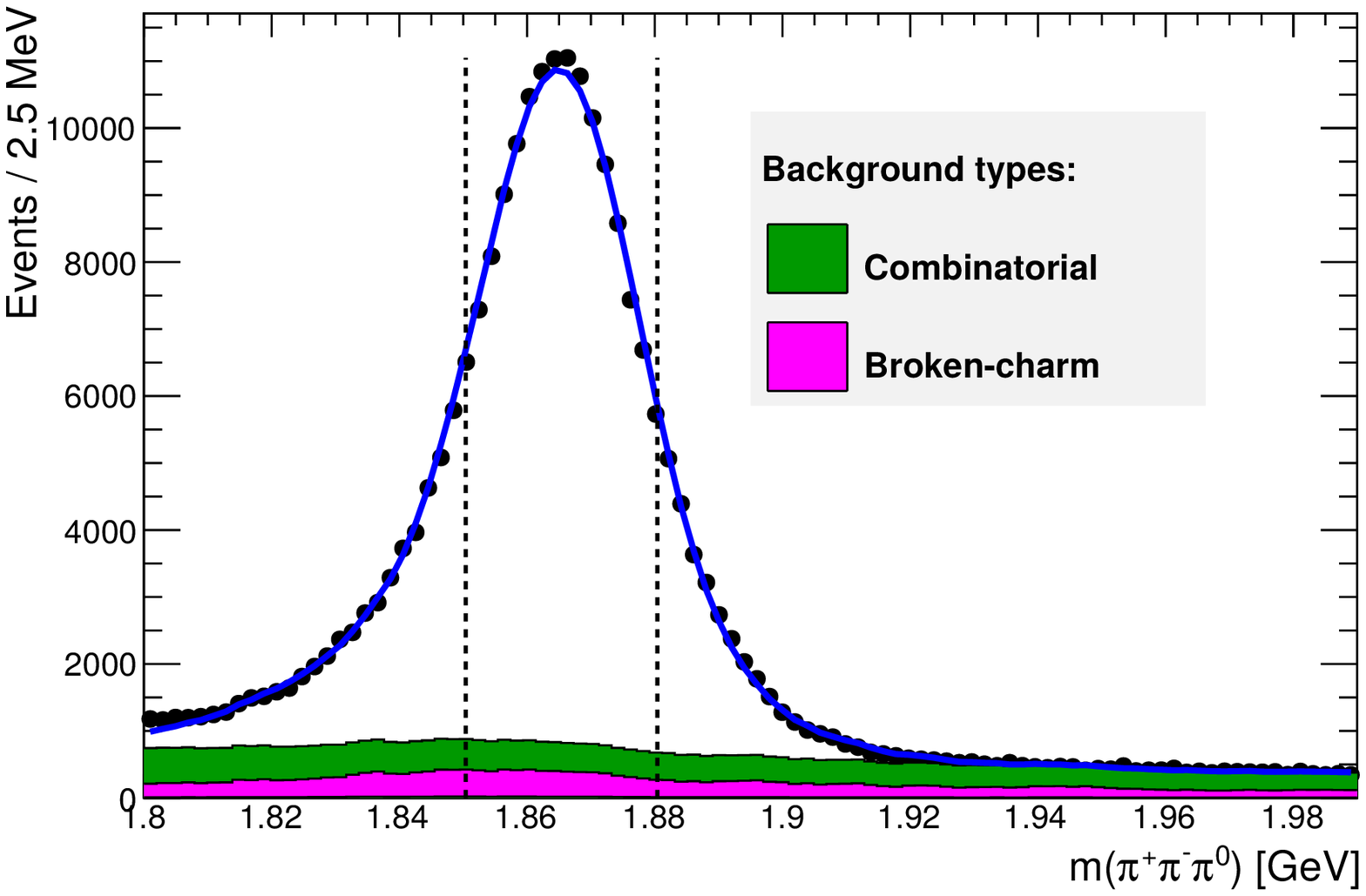}
        \putat{-220}{+120}{ (a)}
    }
    \subfigure{
        \label{fig:massfit-b}
\includegraphics[width=\linewidth]{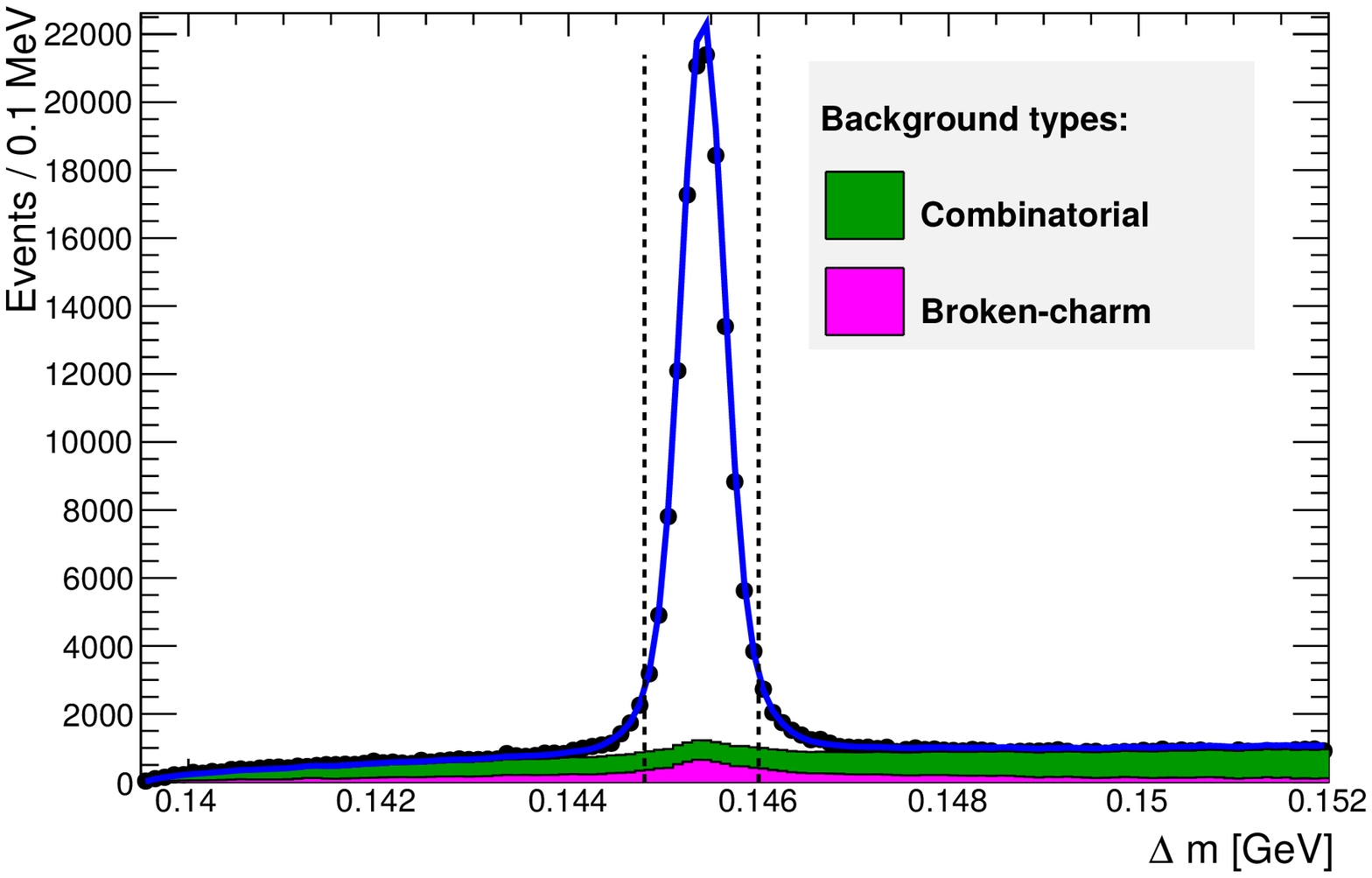}
        \putat{-220}{+120}{ (b)}
    }
    \caption{(Color online). (a) The reconstructed $D^0$ mass distribution of data (dots) with its fit projection (blue line), requiring $\left|\Delta m - \Delta m_{\rm PDG}\right| < 0.6$~\mev; (b) The $\Delta m$ distribution of data (dots) with its fit projection (blue line), requiring $\left| m(\pipi\piz) - m_{\Dz}\right| < 15$~\mev. The underlying histograms shown in shaded bands represent contributions from different background categories defined in Section~\ref{sec:fitstrategy}. 
    The vertical dashed lines mark the actual $m(\pipi\piz)$ or $\Delta m$ requirement for the DP analysis.}
\label{fig:d0massfit}
\end{figure}

Figure~\ref{fig:d0massfit} shows the $m(\pi^+\pi^-\pi^0)$ and $\Delta m$ distributions of $D^0$ candidates passing all the above requirements except for the requirement on the shown variable. 
We relax the requirements on $\Delta m$ and $m(\pi^+\pi^-\pi^0)$ to perform a 
2D-fit in the $m(\pi^+\pi^-\pi^0)$--$\Delta m$ plane, 
whose projections are also shown in Fig.~\ref{fig:d0massfit}. 
The fit determines that about 91\% 
of the $\sim$ 138,000 
candidates satisfying all selection requirements
(those between the dashed lines in Fig. 1), including
those for $m(\pi^+\pi^-\pi^0)$ and $\Delta m$ cuts, are signal.

\section{Measurement of the mixing parameters}
\label{sec:fitstrategy}

\subsection{Fit Model}
\label{sec:fitmodel}
The mixing parameters are extracted through a fit to the DP distribution 
of the selected events as a function of time $t$.
The data is fit with a total PDF which is the sum of 
three component 
PDFs describing the signal, ``broken-charm'' backgrounds, and combinatorial background. 

The signal DP distribution is parametrized in terms of an isobar model~\cite{PhysRev.135.B551,PhysRev.166.1731,PhysRevD.11.3165}.
The total amplitude is a coherent sum of partial 
waves $\mathcal{W}_k$ with complex weights $c_k$,

\begin{equation}
    \overline{A}_f (s_-, s_+) = A_f (s_+, s_-) = \sum_{k}{c_k \mathcal{W}_k(s_+, s_-)}\,,
\label{eq:defamp}
\end{equation}
where $A_f$ and $\overline{A}_f$ are the final state amplitudes introduced in Eq.~(\ref{eq:mixrate}). 
Our model uses relativistic  
Breit-Wigner functions each multiplied by
a real spin-dependent angular factor using
the same formalism with the Zemach 
variation as described in Ref.~\cite{Kopp:2000gv}
for $\mathcal{W}_k$, and 
constant $\mathcal{W}_{\rm{NR}} = 1$ for the 
non-resonant term. 
As in Ref.~\cite{Kopp:2000gv}, $\mathcal{W}_k$ also
includes the Blatt-Weisskopf form factors 
with the radii of $\Dz$ and intermediate resonances
set at 5~$\gev^{-1}$ and 1.5~$\gev^{-1}$, respectively.
The CLEO collaboration modeled the decay as a coherent combination of 
four amplitudes: 
those with intermediate $\rho^{+}, \rho^{0}, \rho^{-}$ 
resonances and a uniform non-resonant term~\cite{PhysRevD.72.031102}. 
This form works well to describe lower statistics samples.
In this analysis we use the model we developed
for our higher statistics search for 
time-integrated $\CP$ violation~\cite{PhysRevLett.99.251801}, 
which also includes other resonances as
listed in Table~\ref{table:fitreson}.
The partial wave with a $\rho^+$ resonance 
is the reference amplitude.
The true decay time distribution at any point in the DP depends
on the amplitude model and the mixing parameters.
We model the observed decay time distribution at each point in the DP
as an exponential with average decay time coming from the
mixing formalism (Eq.~(\ref{eq:mixrate})) convolved with the decay time resolution,
modeled as the sum of three 
Gaussians with  
widths proportional to $\sigma_t$ 
and determined from simulation.
As the
ability to reconstruct $t$ 
varies with the position
in the DP,
our  
parameterization of the signal PDF 
includes $\sigma_t$ functions 
that depend on $ m^2 ( \pi^+ \pi^- ) $,
defined separately in six ranges, 
each as an exponential convolved with a Gaussian.
Efficiency variations across the Dalitz-plot 
are modeled by a histogram obtained from simulated 
decays generated with a uniformly populated phase 
space. 

In addition to correctly reconstructed signal decay chains, 
a small fraction of the events,
$< \, 1$\%, 
contain $D^0\to\pi^+\pi^-\pi^0$ ($\Dzb\to\pipi\piz$) decays
which are correctly reconstructed, 
but then paired with false slow 
pion candidates to create fake $\Dstarp$ ($\Dstarm$) candidates. 
As these are real $ D^0 $ decays, their
DP and decay time distributions are described in the fit
assuming
a randomly tagged flavor. 
The total amplitude 
for this contribution is 
$A'_f(s+, s-) = f_{\rm RS} A_f(s+, s-) + (1-f_{\rm RS})  A_f(s-, s+)$,
where $f_{\rm RS}$ is the ``lucky fraction'' that we have a fake slow pion with 
the correct charge. 
As roughly half of these events are assigned 
the wrong $D$ flavor, we set $f_{\rm RS} = 50\%$ in the nominal fit.
We later vary this fraction to determine a corresponding systematic 
uncertainty.

Backgrounds from mis-reconstructed signal decays and other $\Dz$ decays are referred to as ``broken-charm''. 
In the fit,  the Dalitz-plot distribution for this category is described by histograms taken from the simulations. 
The decay time distributions are described
by the sum of two exponentials convolved with Gaussians
whose parameters are taken from fits to the simulations. 

We use sideband data to estimate combinatorial background.
The data are taken from the sidebands with 
$m\left(\pi^+\pi^-\pi^0\right) < 1.80 \gev$ or $m\left(\pi^+\pi^-\pi^0\right) > 1.92 \gev$, 
and outside of 
the region $0.144 < \Delta m < 0.147$ \gev, 
 where most of the 
broken-charm background events reside. The weighted sum of the two sideband regions is used to 
describe the combinatorial background in the signal region. 
The sideband weights and their 
uncertainties are determined from full MC simulation.
We model these events in $t$ 
similarly to the 
broken-charm category. The decay time is described by the sum of two exponentials convolved 
with Gaussians. As an {\it ad hoc} description of $\sigma_t$ between
0 and 0.8~ps, 
the $\sigma_t$ 
function for the combinatorial background 
is an exponential convolved with a Gaussian, but we 
use different values in six ranges of $\left|t\right|$.

The best-fit parameters are determined by an unbinned maximum-likelihood fit. The central values for 
{\color{black} $x$ and $y$}
were blinded until the systematic uncertainties were estimated. Because of the high  
statistics and the complexity of the model,
the fit is computationally intensive. We have therefore 
developed an open-source framework 
called {\tt{GooFit}}~\cite{goofit}
to exploit the 
parallel processing power of graphical processing units. 
Both the framework and the specific analysis code used in this
analysis are  publicly available 
\footnote{The code is published on GitHub at \url{http://github.com/GooFit/GooFit}}. 

\subsection{Fit Results}
\label{sec:ftiresults}

\begin{figure*}
    \centering
    \subfigure{
        \label{fig:fitres-a}
        \includegraphics[width=0.45\linewidth]{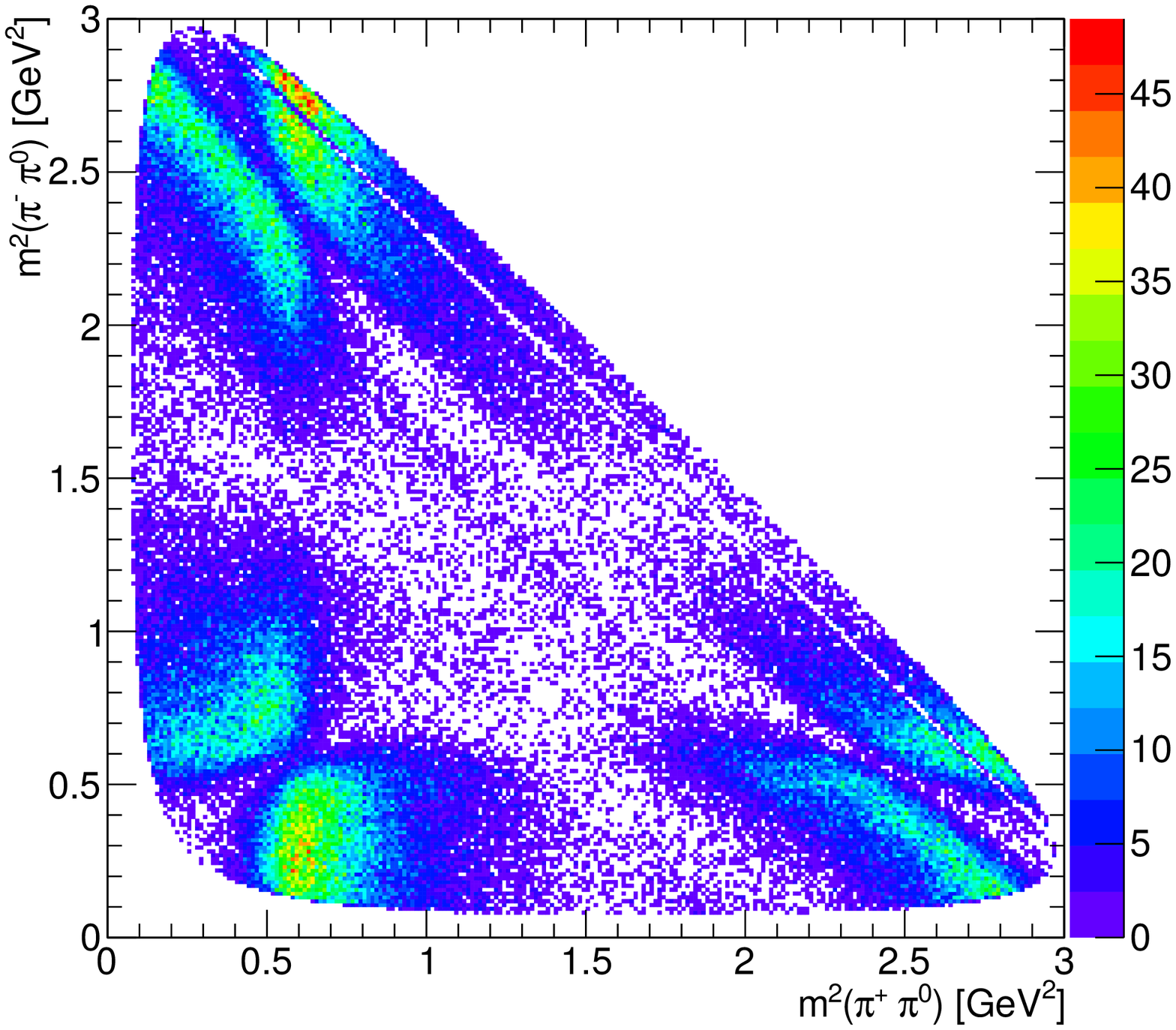}
        \putat{-80}{+170}{ (a)}
    }
    \subfigure{
        \label{fig:fitres-b}
        \includegraphics[width=0.45\linewidth]{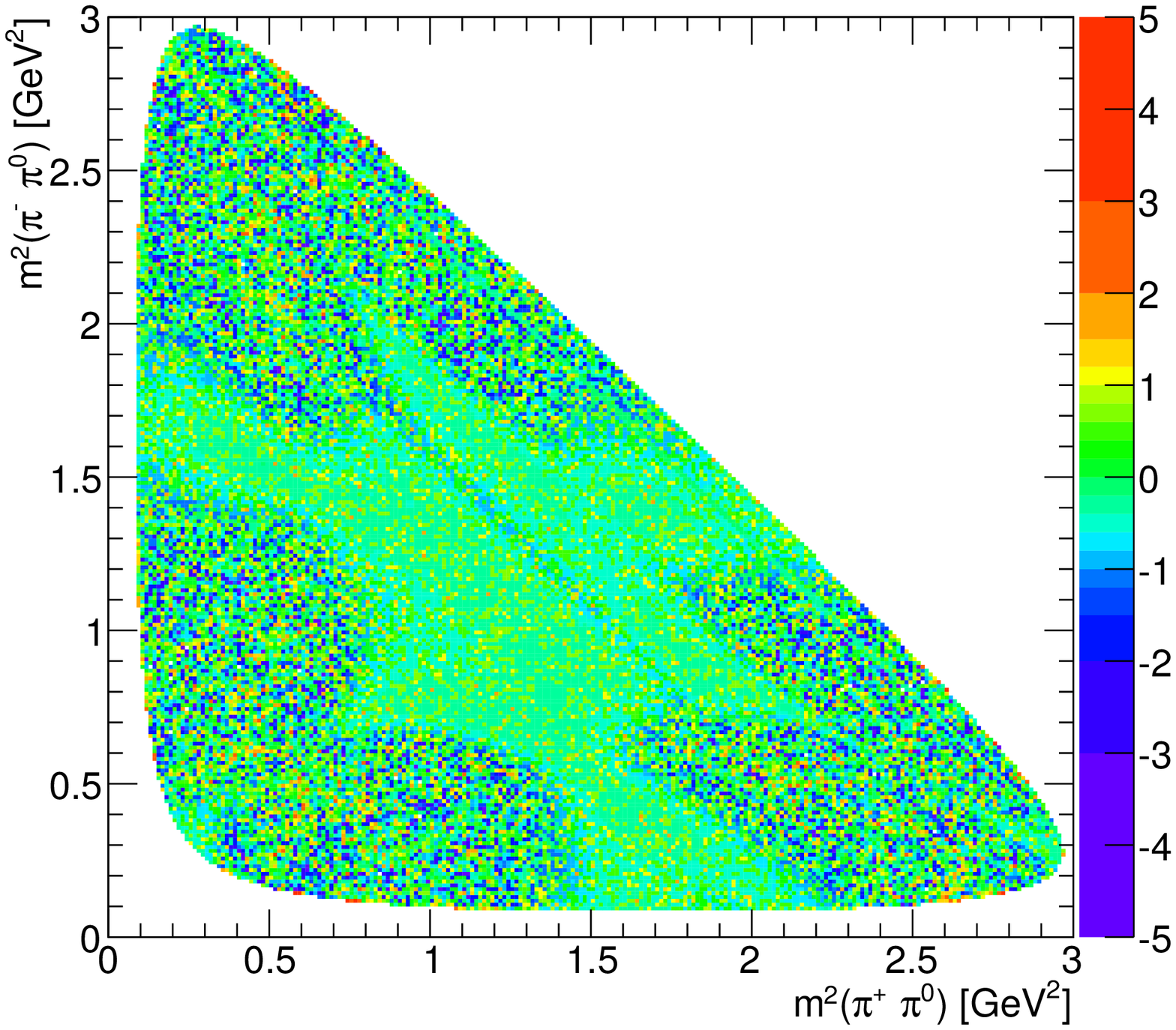}
        \putat{-80}{+170}{ (b)}
    }
    \subfigure{
        \includegraphics[width=0.31\linewidth]{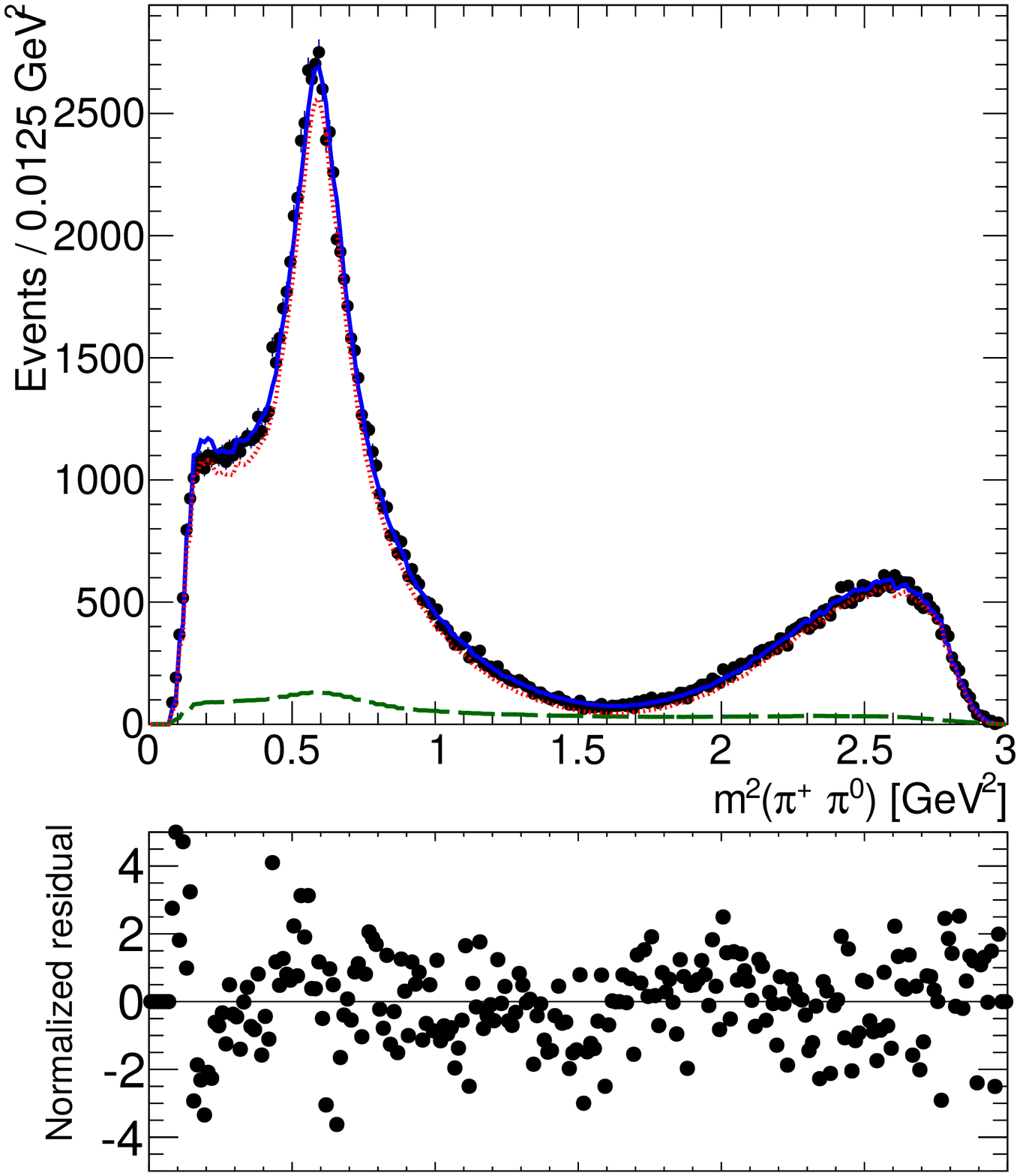}        
        \label{fig:fitres-c}\putat{-50}{+160}{ (c)}
    }
    \subfigure{
        \includegraphics[width=0.31\linewidth]{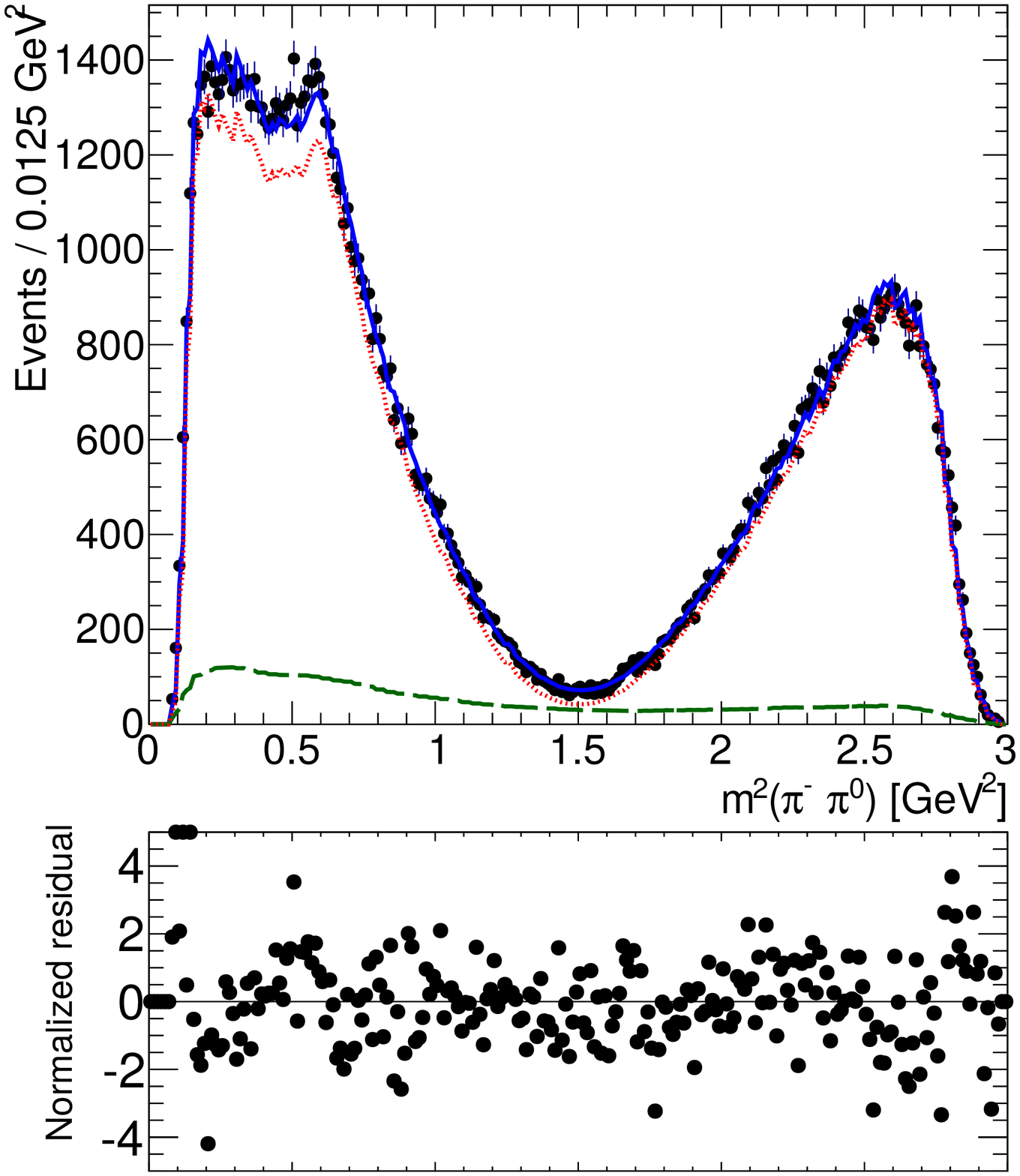}\putat{-50}{+160}{ (d)}
        \label{fig:fitres-d}
    }
    \subfigure{
        \includegraphics[width=0.31\linewidth]{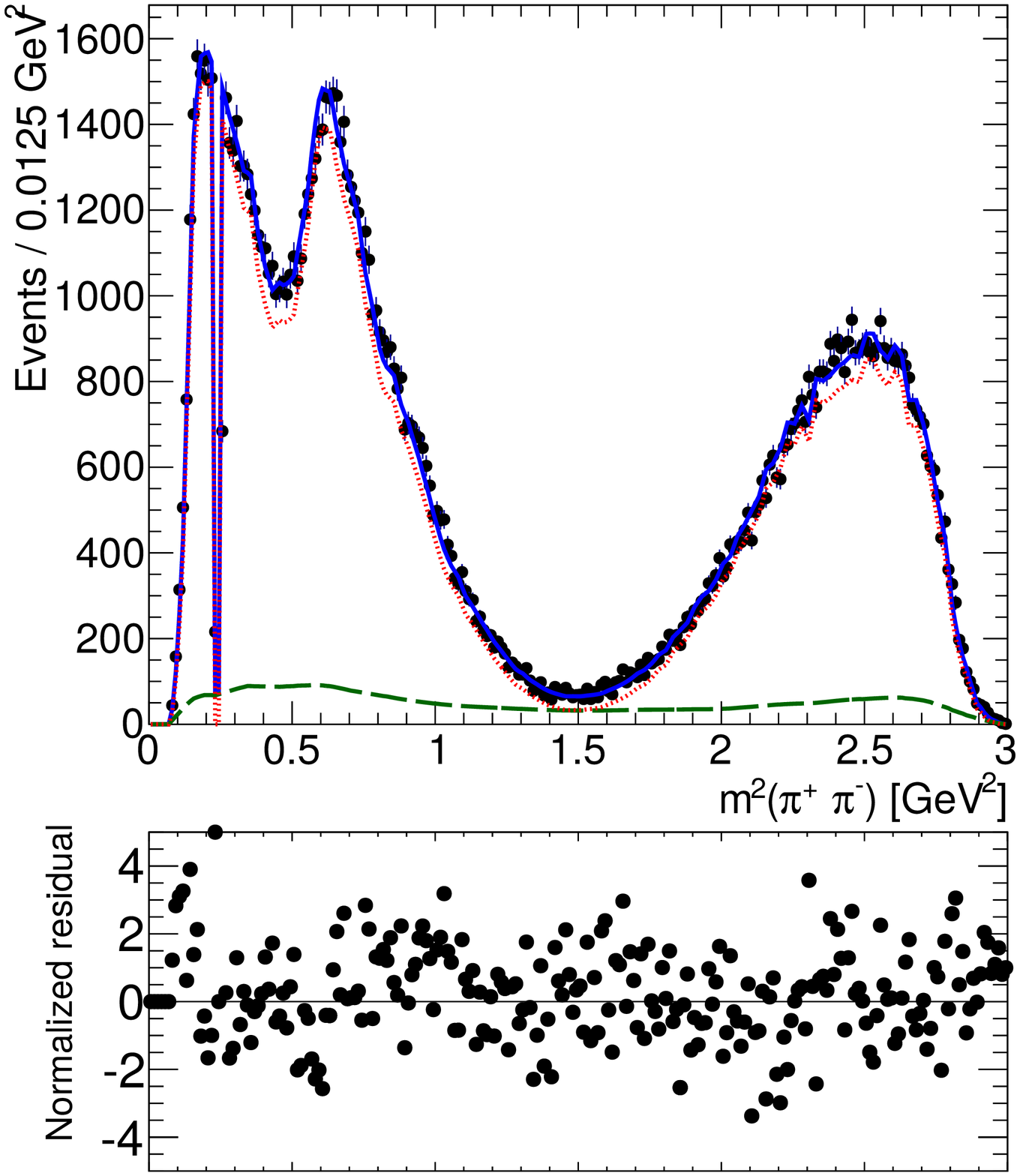}\putat{-50}{+160}{ (e)}
        \label{fig:fitres-e}
    }
    \caption{(Color online). The (a) Dalitz-plot and (b) difference between
    the Dalitz-plot and fit model prediction normalized by the associated statistical uncertainty in each bin, both time-integrated for the data. Also shown underneath are
the projections of (c) $m^2_{\pi^+\pi^0}$, (d) $m^2_{\pi^-\pi^0}$, and (e) $m^2_{\pi^+\pi^-}$
 for our data (points) and fit model (blue solid lines), together with the fit residuals normalized by the associated statistical uncertainties. 
 The PDF components for signal (red dotted) and background (green dashed) events are shown. Note the narrow gap in (e) due to the $\KS$ veto.}
    \label{fig:datadp}
\end{figure*}

The time-integrated Dalitz-plot for the signal region data is
shown in Fig.~\ref{fig:fitres-a}. 
The amplitude parameters determined by the fit described above are listed in
Table~\ref{table:fitreson}. 
Our amplitude parameters
and the associated fractions are generally consistent with the previous 
\babar\ results based on a subset of our data~\cite{PhysRevLett.99.251801}.
The normalized difference between the signal DP and the model 
is shown in Fig.~\ref{fig:fitres-b}. 
The $ m^2 ( \pi^{\pm} \pi^0 ) $ and $ m^2 ( \pi^{+} \pi^{-} ) $ projections of the data and model 
are shown in Fig.~\ref{fig:datadp}(c)--(e).
Differences between the data and the fit model are apparent in
both the Dalitz-plot itself and the projections.
Large pull values are observed
predominantly  near low and high values of $ m^2$ in
all projections.
However, we understand the origin of these discrepancies, and the systematic uncertainties induced on the mixing parameters are small, as discussed below. 
Our fit reports the raw 
mixing parameters as
$x = (\xfitStat) \%$ and 
$y = (\yfitStat) \%$.  
The correlation coefficient between $x$ and $y$  
is $-0.6$\%.
The measured $D^0$ lifetime is $\tau_D = (\tfinalStat){\rm{\,fs}}$,
and agrees with the world average of $(410.1\pm 1.5)$~fs~\cite{pdg}.
The central values of $x$ and $y$ are later
corrected by the estimated fit biases as discussed 
in Sec.~\ref{sec:systematics}.

\begin{table*}
\centering
\caption{\label{table:fitreson}Results of the fit to the $D^0 \to \pi^+\pi^-\pi^0$ sample showing each resonance amplitude magnitude, phase, and fit fraction $f_r \equiv \int{\left|c_k A_k\left(s_+,s_-\right) \right|^2 ds_- ds_+}$. The uncertainties are statistical only. We take the mass (width) of the $f_0(500)$ to be 500 (400) \mev. In the fit, all resonance masses and widths are fixed to the listed values, which are taken from earlier world averages produced by the Particle Data Group~\cite{pdg}. 
} 
\begin{tabular}{l|lcc|,{1.2},{-2.2},{2.1}}
\hline
\hline
               & \multicolumn{3}{c|}{Resonance parameters}
                                                & \multicolumn{3}{c}{ Fit to data results }  \\
State           & $J^{PC}$ & Mass ($\mev$)    & Width ($\mev$)           & \multicolumn{1}{c}{Magnitude} & \multicolumn{1}{c}{Phase ($^\circ$)} & \multicolumn{1}{c}{Fraction $f_r$ (\%)}  \\
\hline
$\rho(770)^+$ & $1^{--}$ & 775.8       & 150.3 & \multicolumn{1}{c}{1}                & \multicolumn{1}{c}{0}            & 66.4,0.5  \\
 $\rho(770)^0$ & $1^{--}$ & 775.8       & 150.3 &  0.55,0.01 & 16.1,0.4&  23.9,0.3 \\
 $\rho(770)^-$ & $1^{--}$ & 775.8       & 150.3  & 0.73,0.01 & -1.6,0.5& 35.6,0.4  \\
 $\rho(1450)^+$& $1^{--}$ & 1465           & 400           & 0.55,0.07 & -7.7,8.2&  1.1,0.3 \\
 $\rho(1450)^0$& $1^{--}$ & 1465           & 400           &   0.19,0.07 & -70.4,15.9& 0.1,0.1\\
 $\rho(1450)^-$& $1^{--}$ & 1465           & 400           &  0.53,0.06 & 8.2,6.7&  1.0,0.2 \\
 $\rho(1700)^+$& $1^{--}$ & 1720           & 250          & 0.91,0.15 & -23.3,10.3& 1.5,0.5 \\
 $\rho(1700)^0$& $1^{--}$ & 1720           & 250          &  0.60,0.13 & -56.3,16.0& 0.7,0.3  \\
 $\rho(1700)^-$& $1^{--}$ & 1720           & 250          & 0.98,0.17 & 78.9,8.5& 1.7,0.6\\
 $f_0(980)$    & $0^{++}$ & 980                 & 44                         &   0.06,0.01 & -58.8,2.9& 0.3,0.1 \\
 $f_0(1370)$   & $0^{++}$ &1434            & 173               &   0.20,0.03 & -19.6,9.5& 0.3,0.1  \\
 $f_0(1500)$   & $0^{++}$ & 1507            & 109           &  0.18,0.02 & 7.4,7.4& 0.3,0.1 \\
 $f_0(1710)$   & $0^{++}$ & 1714            & 140          &  0.40,0.08 & 42.9,8.8& 0.3,0.1  \\
 $f_2(1270)$   & $2^{++}$ & 1275.4        & 185.1 & 0.25,0.01 & 8.8,2.6& 0.9,0.1 \\
 $f_0(500)$    & $0^{++}$ & 500                 & 400                        & 0.26,0.01 & -4.1,3.7& 0.9,0.1 \\
 $NR$          &          &                    &                 &  0.43,0.07 & -22.1,11.7& 0.4,0.1 \\
\hline
\hline
\end{tabular}
\end{table*}


\section{\boldmath Systematic Uncertainties}
\label{sec:systematics}

\begin{table}[!ht]
\centering
\caption{Summary of systematic uncertainties. The various sources are added in 
quadrature to find the total systematic uncertainty.}
\begin{tabular}{lcc}
\hline \hline  \\[-1.7ex] 
Source                              & $x$ [\%] & $y$ [\%] \\ \hline \\[-1.7ex] 
``Lucky'' false slow pion fraction                  &    0.01            &   0.01             \\
Time resolution dependence      &    \multirow{2}{*}{0.03}            &   \multirow{2}{*}{0.02}             \\[-0.5ex]
on reconstructed $D^0$ mass & & \\
Amplitude-model variations&    0.31            &   0.12             \\
Resonance radius                   &    0.02            &   0.10             \\
DP efficiency parametrization  &    0.03            &   0.03             \\
DP normalization granularity &    0.03            &   0.04             \\
Background DP distribution &    0.21            &   0.11             \\
Decay time window        &    0.18            &   0.19             \\
$\sigma_t$ cutoff            &    0.01            &   0.01             \\
Number of $\sigma_t$ ranges             &    0.11            &   0.26             \\
$\sigma_t$ parametrization   &    0.05            &   0.03             \\
Background-model MC time  &    \multirow{2}{*}{0.06}            &   \multirow{2}{*}{0.11}             \\ [-0.5ex]
distribution parameters       & & \\
Fit bias correction                 &    0.29            &   0.02             \\
SVT misalignment                    &    0.20            &   0.23             \\
\hline \\[-1.7ex]
Total                                    &    0.56            &   0.46           \\ \hline \hline
\end{tabular}
\label{table:systematics}
\end{table}

Most  sources of systematic uncertainty are studied by varying some 
aspect of the fit, measuring the resulting $x$ and $y$
values, and taking the full differences
between the nominal and the varied results as the corresponding 
systematic uncertainty.

To study instrumental effects that may not be well-simulated 
and are not covered in other studies,
we divide the data into four groups of disjoint bins and 
calculate $ \chi^2 $ with respect to the overall average
for each group for both $ x $ and $ y $.
Within a group, each bin has roughly the same statistics.
Four bins of $ m ( \pipi \pi^0 ) $
give $ \chi^2 = 3.9 $ (0.2) for $ x $ ($ y $); 
five bins of each of $\Dz$ laboratory momentum $ p_{\rm lab}$, $ \cos \theta $,
and $ \phi $ 
give $ \chi^2 $ values of 1.5, 1.2, and 3.2
(5.9, 5.1, and 6.9) for $ x $ ($ y $), respectively.
Altogether, the summed $ \chi^2 $ is 27.9 for $ \nu = 37 $ degrees of
freedom.
Ignoring possible correlations, the $ p $-value for the 
hypothesis that the variations are consistent with 
being purely statistical fluctuations around a common
mean value is $ \approx 85 \% $.
Therefore, we assign no additional systematic
uncertainties.

Table~\ref{table:systematics} summarizes the systematic uncertainties
described in detail below.
Combining them in quadrature, we find total systematic uncertainties
of 0.56\% for $x$ and 0.46\% for $y$.

As mentioned earlier, one source of background comes from
events in which the $D^0$ is correctly reconstructed, but is paired 
with a random slow pion. 
We assume the lucky fraction $f_{\rm RS}$ to be exactly $50\%$ in the nominal fit. 
To 
estimate the uncertainty 
associated with this assumption,
we vary the fraction from 40\% to 60\% and take 
the largest variations as an estimate of the uncertainty.

The detector resolution leads to correlations between reconstructed
$ D^0 $ mass and the decay time, $ t $.
We 
divide the sample into four ranges of $ D^0 $ mass with approximately
equal statistics
and fit them separately; 
we find the variations consistent with statistical fluctuations. 
Because the average decay time is correlated with 
the reconstructed $ D^0 $ mass,
we refit the data by introducing separate time resolution functions 
for each range, allowing the sets of parameters to vary independently. 
The associated systematic uncertainties are taken as the differences from 
the nominal values.

The DP distribution of the signal is modeled as a coherent sum of quasi-two-body decays, involving 
several resonances. 
To study the sensitivity to the choice of the model, we remove some
resonances from the coherent sum. 
To decide if removing a resonance provides 
a ``reasonable'' description of the data,
we  calculate the $\chi^2$ of a fit using an adaptive binning  process
where each bin contains at least 
a reasonable number of events so that its statistical
uncertainty is well determined. 
With 1762 bins, the nominal fit has $ \chi^2 = 2794 $.
We separately drop the four partial waves that individually
increase $ \chi^2 $ by less 
than 80 units: 
$f_0(1370)$, $f_0(1500)$, $f_0(1710)$, and $\rho(1700)$. 
We take the largest variations as the systematic 
uncertainties. 
The other partial waves 
individually when removed produce $ \Delta \chi^2 > $ 165.
Additional uncertainties from our amplitude model due to poor knowledge
of the mass and width of $f_0(500)$ are accounted for
by floating the mass and width of $f_0(500)$ in the fit to data and taking 
the variations in $x$ and $y$.
The default resonance radius used in the Breit-Wigner resonances 
in the isobar components 
is 1.5~$\gev^{-1}$, as mentioned earlier. 
We vary it in steps of 0.5 $\gev^{-1}$ from a radius of 0 to 2.5~$\gev^{-1}$
and again take the largest 
variations.

The efficiency as a function of position in the DP in the 
nominal fit is modeled using a histogram taken
from events generated with a uniform phase space distribution. 
As a variation, we parameterize the 
efficiency using a third-degree polynomial in $s_+, s_-$ and take the 
difference in mixing parameters 
as the uncertainty in the efficiency model. 
Normalization over the DP is done numerically by evaluating 
the total PDF on a
$120 \times 120$
grid. 
To find the sensitivity to the accuracy of the normalization integral, we 
vary the granularity of the grid from  $ 120 \times 120 $  to 
$240 \times 240$ 
 and 
take the largest variations as systematic uncertainties.
The 
 combinatorial background in the DP is modeled by sideband data summed according to weights taken 
 from simulation. 
We repeat the fit using a histogram taken from simulation and vary 
 the weights by $\pm 1$ standard deviation. 
Additionally, we vary the number of bins used in the 
 ``broken-charm'' histograms.

In the nominal fit, we consider events in the decay time window 
between $-2$ and $+3$~ps, $\ie$ about $-5$ to $+7$~$\tau_{\Dz}$.
To test our sensitivity to high-$| t  |$ events, 
the window is varied, with the low end ranging from $-$3.0 ps to $-$1.5 ps
and the high end ranging from 2.0 ps to 3.0 ps.
 We assign an 
 uncertainty of 0.18\% to $x$ and 0.19\% to $y$, 
the 
largest variations from this source. 
We vary the maximum allowed uncertainty on the 
reconstructed decay time $\sigma_t$ to study the effect of 
poorly measured events. The nominal 
cutoff at 0.8 ps is relaxed to 1.2 ps in steps of 0.1 ps and we use the largest variations as the uncertainties 
from this source. To account for the variation of $\sigma_t$ across the DP,
 the nominal fit has six different 
$\sigma_t$ distributions, one for each range of $m^2(\pi^+\pi^-)$. We reduce the number of ranges to two 
and increase it to eight, and use the largest difference as the uncertainty associated with the number of ranges. 
Additionally, instead of using a functional form to describe the $\sigma_t$ distribution in each range, we 
repeat our nominal fit using a histogram taken from simulation. This produces extremely small changes in the 
measured mixing parameters; we take the full difference as an estimate of the uncertainty.

In the nominal fit, the background components have their 
decay time dependences modeled by the sums of two 
exponentials convolved with Gaussians whose parameters are fixed to values found from fits to simulated data. 
We vary each parameter in sequence by $\pm 1$ standard deviation and take the largest 
variations as estimates of the systematic uncertainty.

Our fits 
combine 
two effects: detector resolution and efficiency. 
We ignore the migration of
events which are produced at one point in the DP and reconstructed at another point;
we parameterize
detection efficiency from simulated events, generated with a uniformly
populated DP using the observed positions, in the numerator.
As noted earlier, this leads to discrepancies between fit projections and
data for simulated data which are very similar to those observed for 
real data as observed in
Fig.~\ref{fig:datadp}. 
We believe this is due to ignoring the systematic migration of events away
from the boundaries of phase space induced by misreconstruction followed
by constrained fitting.
We have further checked the migration effect by fitting the data in a 
smaller DP phase space with all the boundaries shifted 0.05 $\gev^2$ inwards.
In addition, detector resolution leads to a correlation between reconstructed $ D^0 $ mass
and $ t $, also noted earlier.
To estimate the level of bias and systematic uncertainty introduced by 
these factors, we studied the full MC samples described in 
Section~\ref{sec:evtsel}.
The fit results display small biases in $x$ and 
$y$. 
{\color{black}From the fit to each sample, we determine the
    pull values for $x$ and $y$, defined as the differences
       of fitted and input values. 
We then correct for  fit biases by subtracting $+0.58\%$ from $x$ 
and $-0.05$\% from $y$ where the numerical values are the 
mean deviations from the generated values.
}
The assigned systematic uncertainties are
half the shifts in each variable.

To test the sensitivity of our results to small uncertainties in our knowledge of the 
precise positions of  
the SVT wafers, we reconstruct some of our MC samples with 
deliberately wrong alignment files that produce much greater pathologies
than are evident in the data. 
We again create background mixtures and fit these misaligned samples. 
Four samples are generated, all with $x= y=+1\%$. 
Each 
sample 
has roughly the same magnitude of effect caused 
by the five different 
misalignments considered. 
As the misalignments used in this study are extreme,
we estimate the
systematic uncertainties as half of the averages of the absolute values of the 
shifts in $x$ and $y$.


\section{Summary and conclusions}
\label{sec:conclusion}

We have presented the first measurement of $ \Dz $--$\Dzb$
mixing parameters from a time-dependent amplitude analysis of
the decay $ D^0 \to \pipi \pi^0 $.
We find $ x = (\xfinalf) \% $ and $ y = (\yfinalf) \%$, 
where the
quoted uncertainties are statistical and systematic, respectively.
The dominant sources of systematic  uncertainty can be reduced
in analyses with larger data sets.
Major sources of systematic uncertainty in this measurement
include those 
originating in how we determine shifts for detector misalignment
and the choice of decay time window.
We estimated conservatively 
the former as it is already small compared to
the statistical uncertainty of this measurement.
The latter can be reduced by more carefully determining the 
signal-to-background ratio as a function of decay time.
However, since the 
systematic uncertainties are already small compared to the statistical uncertainties, we choose not to do so in this analysis.
Similar considerations suggest that systematic uncertainties
will remain smaller than statistical uncertainties even when data
sets grow to be 10 to 100 times larger in experiments such
as LHCb and Belle II.


\section{Acknowledgments}
We are grateful for the 
extraordinary contributions of our \pep2\ colleagues in
achieving the excellent luminosity and machine conditions
that have made this work possible.
The success of this project also relies critically on the 
expertise and dedication of the computing organizations that 
support \babar.
The collaborating institutions wish to thank 
SLAC for its support and the kind hospitality extended to them. 
This work is supported by the
US Department of Energy
and National Science Foundation, the
Natural Sciences and Engineering Research Council (Canada),
the Commissariat \`a l'Energie Atomique and
Institut National de Physique Nucl\'eaire et de Physique des Particules
(France), the
Bundesministerium f\"ur Bildung und Forschung and
Deutsche Forschungsgemeinschaft
(Germany), the
Istituto Nazionale di Fisica Nucleare (Italy),
the Foundation for Fundamental Research on Matter (The Netherlands),
the Research Council of Norway, the
Ministry of Education and Science of the Russian Federation, 
Ministerio de Econom\'{\i}a y Competitividad (Spain), the
Science and Technology Facilities Council (United Kingdom),
and the Binational Science Foundation (U.S.-Israel).
Individuals have received support from 
the Marie-Curie IEF program (European Union) and the A. P. Sloan Foundation (USA). 


\bibliography{refs_bad_pipipi0}

\end{document}